\documentclass[12pt]{article}
\usepackage{epsfig}
\usepackage{psfig}
\begin{document}
\begin{center}
Motion of Massive Particles in Rindler Space and the Problem of
Fall at the Centre

\medskip
Sanchita Das$^a$, Soma Mitra$^b$ and Somenath Chakrabarty$^c$ \\
Department of Physics, Visva-Bharati, Santiniketan, India 731235 \\
$^a$Email:sanchita.vbphys@gmail.com\\
$^b$Email:somaphysics@gmail.com\\
$^c$Email: somenath.chakraborty@visva-bharati.ac.in
\end{center}

\medskip
\begin{center}
Abstract
\end{center}
The motion of a massive particle in Rindler space has been studied 
and obtained the geodesics of motion. The orbits in Rindler space are
found to be quite different from that of Schwarzschild case. The paths
are not like the Perihelion Precession type. Further we have
set up the non-relativistic Schrodinger equation for the particle in the
quantum mechanical scenario in presence of background constant gravitational field and
investigated the problem of fall of the particle at the center.  This
problem is also treated classically. Unlike the conventional scenario,
here the fall occurs at the surface of a sphere of unit radius.

\bigskip
It is well known that 
the Lorentz transformations of space time coordinates are
between the inertial frame of
references \cite{R1}, whereas the Rindler transformations are between an inertial 
frame and 
the frame having uniform accelerated motion \cite{R2,R3,R4,R5,R6,R7,R8,R8a,R9}. The space here is called 
the Rindler space, exactly like the Minkowski space in the Lorentz 
transformation scenario. Further, the Rindler space is locally flat,
whereas the flatness of the
Minkowski space is global in nature. 
Now according to the  principle of equivalence a frame undergoing 
accelerated motion in absence of gravity is equivalent to a frame at rest 
in presence of a gravitational field. The strength of the gravitational 
field is exactly 
equal to the magnitude of the acceleration of the moving frame. Therefore the Rindler  space 
may be considered to be a space associated with a frame at rest in presence 
of a 
uniform gravitational field. The Rindler space-time  transformations
in natural unit ($c=\hbar =1$) and in $1+1$  dimension are given by \cite{R7,R8,R8a,R9}
\begin{equation}
t=\left(\frac{1}{\alpha}+x^{'}\right) \sinh(\alpha t^{'})
\end{equation}
\begin{equation}
x=\left(\frac{1}{\alpha}+x^{'}\right) \cosh(\alpha t^{'})
\end{equation}
Where primed coordinates are in the 
non-inertial frame, whereas  the unprimed  are 
in  the inertial one. The Rindler space-time  transformations are therefore 
exactly like the Lorentz  transformations but in uniformly accelerated frame. Here 
$\alpha$ is the local acceleration or local uniform gravitational field.
Hence the line element in Rindler space in $1+1$-dimension is given by
\begin{equation}     
ds^2=-d\tau^2=-(1+\alpha x)^2dt^2+dx^2
\end{equation}
where $\tau$ is the proper time. Then one can  define the classical action integral in the form \cite{R1}
\begin{equation} 
\int_A^B d\tau=\int_A^B L d\lambda
\end{equation}
 where $L=\frac{d\tau}{d\lambda}$,
 the Lagrangian of the particle, given by \cite{R10,R11,R12}
  \begin{equation}
 L=\left [(1+\alpha x)^2\left (\frac{dt}{d\lambda}\right )^2-\left
 (\frac{dx}{d\lambda}\right)^2\right ]^\frac{1}{2}
  \end{equation}
The aim of the present article is to study the particle motion of non-zero mass in Rindler
space and obtain the geodesics of motion. We further investigate the problem of fall at the
origin. The last one is based on both the classical motion and the quantum mechanical motion of the
particle. The particlles will occupy the space near the surface of the sphere of radius unity.

Now the well known Euler-Lagrange equation is given by
  \begin{equation} 
 \frac{\partial}{\partial \lambda}\left (\frac{\partial L}{\partial(\frac{\partial
 q_i}{\partial \lambda})}\right )-\frac{\partial L}{\partial q_i}=0
 \end{equation}
 where $q_i=t$ or $x$. Let us first consider $q_i=t$, the time coordinate. Since $t$ is
 a cyclic coordinate, $\frac{\partial L}{\partial t}=0$ and therefore
 \begin{equation}
 \frac{\partial t}{\partial \tau}=\frac{C}{(1+\alpha x)^2}
  \end{equation}
 where $C$ is a constant and we have used the expression for $L$  as given by eqn.(5).
We next consider $q_i=x$, then with some little algebra, we have from the Euler-Lagrange equation
\begin{equation} 
 (1+\alpha x)\alpha \left (\frac{\partial t}{\partial \tau}\right )^2+\frac{d^2x}{d\tau ^2}=0
   \end{equation}
 On substituting the expression for $\frac{\partial t}{\partial \tau}$ from eqn.(7) we have
 \begin{equation} 
 \frac{\alpha C}{(1+\alpha x)^3}+\frac{d^2 x}{d\tau ^2}=0
 \end{equation}
(see Appendix). Now changing the variable from $x$ to $u=1+\alpha x$ and redefining 
$\tau\rightarrow C^\frac{1}{2}\alpha \tau$, we have from the above equation,
 \begin{equation} 
 \frac{d^2 u}{d \tau^2}+\frac{1}{u^3}=0
 \end{equation} 
This is the equation of motion for the particle in Rindler space.
Integrating both the sides with respect to $u$, this equation may be transformed to the form
 \begin{equation} 
 \frac{1}{2}\left (\frac{du}{d\tau}\right )^2-\frac{1}{2u^2}={\rm{constant}}
 \end{equation}
Rewriting $\tau$ in its original form and defining $p=du/d\tau$ as the momentum of the particle of
unit mass, we have from the above equation
\begin{equation}
\frac{p^2}{2}-\frac{C\alpha^2}{2u^2}={\rm{constant}}=E
\end{equation}
where the first term on the left hand side is like the kinetic energy for a unit mass and
the second term is an attractive inverse square potential, whereas the constant on the right 
hand side may be treated as the total energy of the particle. We further assume that the
energy of the particle is negative in nature, i.e., we are considering bound orbit of the
classical particle. Now expressing the above equation  in the usual form, given by
  \begin{equation} 
 \frac{1}{2} p^2+V(u)=E
 \end{equation}
where $V(u)=-\frac{C\alpha^2}{2u^2}$, the inverse square attractive potential, then depending on the
strength $0.5 C\alpha^2$ of the attractive potential $V(u)$, the particle may fall at the centre
($x=0$), where the momentum becomes zero ($p=0$, i.e., particle is at rest) and in this
situation $E=V(u=1)=-0.5 C\alpha^2$, the maximum depth of the potential, which is also the
minimum possible value of the total energy. Since it is negative it gives the maximum binding
at $x=0$ or $u=1$.  In fig.(1) we have shown this potential well. The coordinate $x$ is
plotted along $x$-axis, whereas the magnitude of the potential $\mid V(u)\mid$ is
plotted along $y$-axis. Therefore in actual scenario this is not the potential hill but is
potential well. But unlike the conventional scenario of fall at the centre problem, there is
no singularity at the origin ($x=0$), or at the centre of the gravitating object \cite{R13,R14,R15}. The value
of $V(u)$ is $-0.5C\alpha^2$ at the origin and $\longrightarrow -1$ for large $u$ values.
Therefore unlike the conventional case, where the potential part $\longrightarrow -\infty$ at the origin, here, 
since the minimum value of $u$ is positive unity, therefore $V(u)\rightarrow -0.5C\alpha^2$ for
$x\longrightarrow 0$ or $u\longrightarrow 1$. 
In this case the singularity is as if covered by a sphere of unit radius. Therefore here the problem of fall at 
the origin is reduced to the problem of fall on the unit surface. 

For further investigation of this problem of fall at the centre, to some extent elaborately, 
let us  get the geodesics of motion for the particle in Rindler space.
We start with the equation
\begin{equation} 
(1+\alpha x)^2\left (\frac{dt}{d\tau}\right )^2-\left (\frac{dx}{d\tau}\right )^2=1 
\end{equation}
Hence we have after substituting for $\frac{dt}{d\tau}$ from eqn.(7)
\begin{equation}
\tau=\large\int_0^x \frac{dx}{[\frac{C}{(1+\alpha x)^2}-1]^\frac{1}{2}} 
\end{equation}
Using the changed variable $u(=1+\alpha x)$, we have,
\begin{equation}
\alpha \tau=-\frac{C^{1/2}}{\alpha}\left[ \left\lbrace 1-\left (\frac{1+\alpha
x}{C^{1/2}}\right )^2 \right\rbrace^\frac{1}{2}-\left(1-\frac{1}{C}\right)^\frac{1}{2}\right]
\end{equation}
Since the uniform acceleration $\alpha$ of the frame is completely arbitrary, we shall study
the variation of $\alpha x$ with $\alpha \tau$. The above equation can also be expressed
in the following more convenient form
\begin{equation}
\alpha x=C^{1/2} \left [1-\left \{\left (1-\frac{1}{C}\right )^{1/2}-\frac{\alpha
\tau}{C^{1/2}}\right \}^2\right ]-1
\end{equation}
which will give the variation of $\alpha x$ with $\alpha \tau$. Of course $\alpha$ is
assumed to be constant throughout, whereas by evaluating $\alpha x$ as a function of $\alpha
\tau$, we have removed the arbitrariness of $\alpha$.

Before we obtain numerically  the variation of $\alpha x$ with $\alpha \tau$, we shall impose 
some constraints on the quantities appearing in the above equation. By inspection one 
can put the following restrictions: $C> 1$, and if $\alpha x> 0$, then $\alpha \tau > 0$. 
Since we are  measuring $x$ along positive direction and the frame is also moving with
uniform acceleration $\alpha $ along positive $x$-direction, the second constraint is
is also quite obvious. In fig.(2) we have shown the variation of $\alpha x$ with $\alpha \tau$ for
three different $C$ values. From the curves one may infer that the geodesics are closed in
nature; returned to the initial point, i.e., at $ x\approx 0$ (Boomerang type geodesic). The geodesics are
not like the Perihelion Precession type as has been obtained in the Schwarzschild geometry \cite{R16,R17}.
For the proper justification of our arguments as given above, we
next consider the equation of motion of the particle under inverse square potential, given 
by eqn.(10). This equation has been solved numerically for the initial conditions
$\frac{du}{d\tau}=0$, i.e., the motion is assumed to be started from rest and for a number of initial
positional coordinates in the scaled form given by $u_0$. 
The boundary value of $u$ is unity, i.e., on the surface of a unit sphere.
In fig.(3) we have shown the geodesics of motion for three different $u_0$ values.

Now to complete the study of free fall at the centre in our modified formalism, let us next 
investigate the quantum mechanical motion of the particle in inverse square 
attractive potential \cite{R13,R14,R15}. We consider the bound state problem, i.e., $E<0$. Re-substituting the 
exact form of $\tau$, we have the eigen value equation for a unit mass particle
\begin{equation}
H\psi(u)=(\frac{p^2}{2}-\frac{\beta}{u^2})\psi(u)=-E\psi(u)
\end{equation}
where $\beta=\frac{\alpha^2C}{2}$.
Hence we have
\begin{equation} 
\frac{d^2\psi}{du^2}+\frac{\gamma}{u^2}\psi(u)-K^2\psi(u)=0
\end{equation}
where $\gamma=\frac{2\beta}{\hbar^2}$ and $K^2=\frac{2E}{\hbar^2}$.
Therefore in the asymptotic region, i.e., $u\rightarrow \infty$, $\psi\sim e^{-Ku}$ or it goes
to zero. The wave function is therefore well behaved at infinity. Next to study the nature of wave function near the unit surface, we substitute
$\psi(u)=\frac{R(u)}{u}$.
Then it can very easily be been shown that $R(u)$ satisfies the equation
\begin{equation}
u^2\frac{d^2R}{du^2}-2\rho\frac{dR}{du}+(\gamma-2-K^2 u^2)R=0
\end{equation}
Putting $Ku=\rho$ as a new variable, we have,
\begin{equation}
\frac{d^2R}{d\rho^2}-\frac{2}{\rho}\frac{dR}{d\rho}+\frac{\gamma-2}{\rho ^2}R-R=0
\end{equation}
To investigate the nature of $R$ near unit surface we write $\rho=\rho_0$ in the denominator 
of the second and the third term and then put the limit $\rho_0\rightarrow 1$.
Then we have
\begin{equation}
\frac{d^2R}{d\rho^2}-2\frac{dR}{d\rho}+(\gamma-3)R=0
\end{equation}
Then for $\gamma<4$, we have the solution
\begin{equation}
u(\rho)=\exp(\rho)[A_1 \exp((4-\gamma)^{1/2}\rho)+A_2
\exp(-(4-\gamma)^{1/2}\rho)]
\end{equation}
whereas for $\gamma>4$, the solution is given by
\begin{equation}
u(\rho)=\exp(\rho)[A_1 \exp\{i(\gamma -4)^{1/2}\rho\}+
A_2\exp\{-i(\gamma -4)^{1/2}\rho\}]
\end{equation}
The solutions are of standard stationary type and exist for $\rho=\rho_0\rightarrow 1$.
Further, for bound state solution, i.e., for $E<0$, $E_k<|V|$, where $E_k$ and $V$ are the 
kinetic energy and attractive potential respectively. The quantity $|V|$ is maximum for 
$\rho=\rho_0=K$, i.e., at $x=0$ and  approaches unity as $\rho$ increases to infinity. 
Therefore the negative value of $E$ is maximum on the surface of the unit shell. This 
indirectly indicates the "fall" on the surface. The basic difference with the usual solution is 
that the fall is not at the centre $(\rho=0)$ which is covered by the unit shell.
Therefore we may conclude that in the Rindler space the energy of the particle will minimum
on the surface of the sphere of unit radius. This indirectly indicate that all the particles
will fall on this surface because of the minimum values of the energy of the particles. This
is true for classical as well as quantum mechanical scenarios. The other interesting finding
of this work is that unlike the perihelion precession type geodesics these are closed type.
We call them as boomerang geodesics.
\begin{center}
Appendix
\end{center}
In this appendix we shall give a short outline to show that 
\begin{eqnarray}
&&\frac{1}{L}\left [(1+\alpha x)\alpha \left (\frac{dt}{d\lambda}\right )^2+\frac{d^2x}{d\lambda
^2}\right ]\nonumber \\
&=&\frac{1}{L}\left [(1+\alpha x)\alpha \left (\frac{dt}{d\tau}\right )^2
\left (\frac{d\tau}{d\lambda}\right )^2+\frac{d^2x}{d\tau^2}\frac{d^2\tau}{d\lambda
^2}\right ]
\nonumber \\
&=&\frac{\alpha C}{(1+\alpha x)^3}+\frac{d^2x}{d\tau ^2}=0
\end{eqnarray}
we have
\begin{equation}
\frac{dt}{d\lambda}=\frac{dt}{d\tau}\frac{d\tau}{d\lambda}=L\frac{dt}{d\tau}
\end{equation}
Hence
\begin{equation}
 \left (\frac{dt}{d\tau}\right )^2=L^2\left (\frac{dt}{d\tau}\right )^2
 \end{equation}
\begin{equation} 
\frac{d^2 x}{d\lambda ^2}=\frac{d}{d\lambda}\left (\frac{dx}{d\lambda}\right )
=\frac{d}{d\tau}\left (\frac{dx}{d\lambda}\right )\left (\frac{d\tau}{d\lambda}\right )
\end{equation}
\begin{equation} 
\frac{d^2 x}{d\lambda ^2}=\frac{d}{d\lambda}\left (\frac{dx}{d\tau}\right )
 \left (\frac{d\tau}{d\lambda}\right )
\end{equation}
\begin{equation} 
\frac{d^2 x}{d\lambda ^2}=\frac{d^2x}{d\tau ^2}\left (\frac{d\tau}{d\lambda}\right )^2
=L^2\frac{d^2x}{d\lambda ^2}
\end{equation}
Hence we have the above equality after putting, $\frac{dt}{d\lambda}=\frac{C}{(1+\alpha x)^2}$

\newpage
\begin{figure}[ht]
\psfig{figure=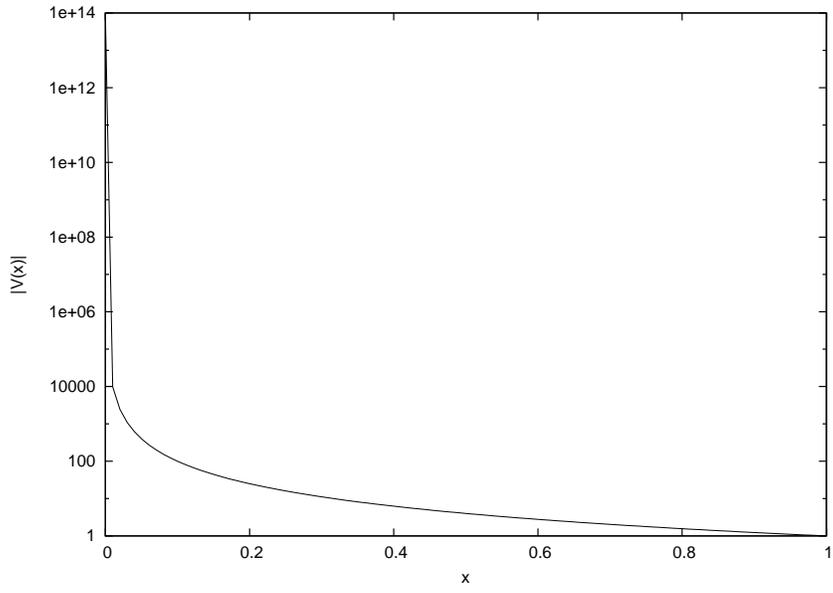,height=0.8\linewidth,angle=-90}
\caption{The variation of $\mid V(x)\mid$ with $x$}
\end{figure}
\begin{figure}[ht]
\psfig{figure=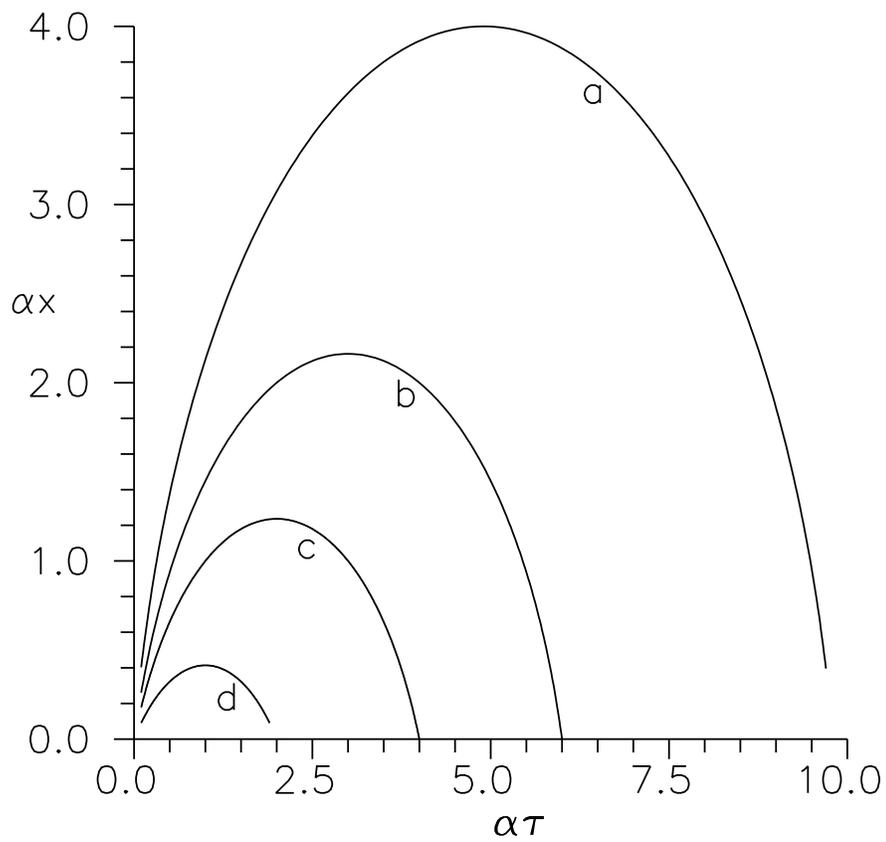,height=0.8\linewidth,angle=0}
\caption{The geodesics of motion for various $C$ values: curve (a) is for $C=25$, (b) is for 
$C=10$, (c) is for $C=5$ and (d) is for $C=2$.} 
\end{figure}
\begin{figure}[ht]
\psfig{figure=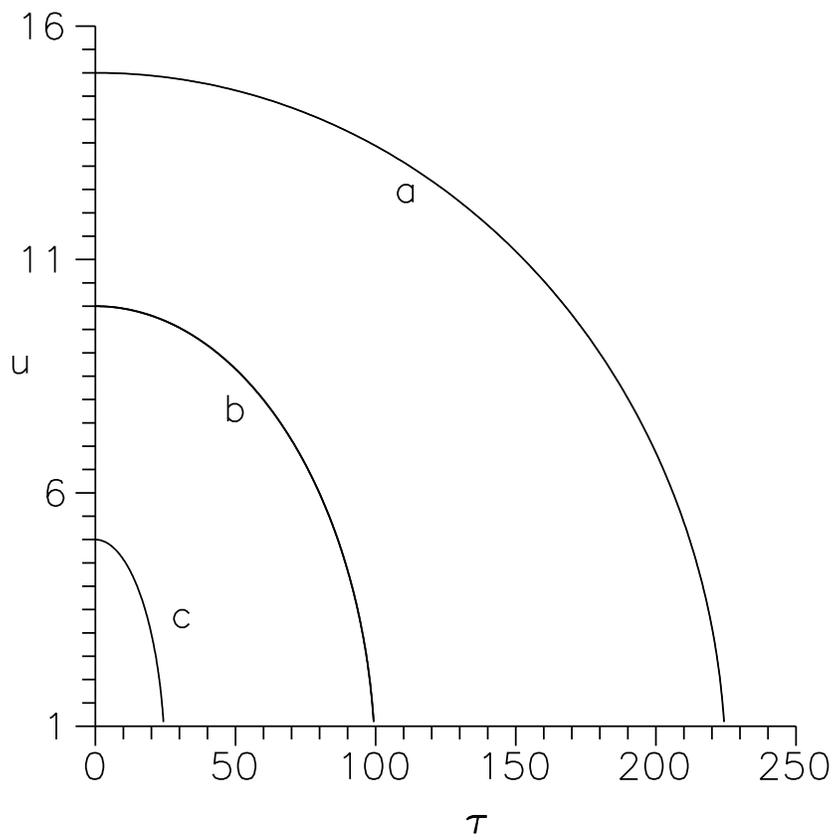,height=0.8\linewidth,angle=0}
\caption{The geodesics of motion for various $u_0$ values: curve (a) is for $u_0=15$, (b) is for $u_0=10$ and (c) is
for $u_0=5$}
\end{figure}
\end{document}